# Electron and hole trapping in the $Ce^{3+}$ and $Pr^{3+}$ doped lutetium pyrosilicate scintillator crystals studied by electron paramagnetic resonance

V. Laguta,[1] M. Buryi,[1] Y. Wu,[2] G. Ren,[2] M. Nikl[1]

[1]*Institute of Physics CAS Prague, Cukrovarnicka 10/112, Prague, Czech Republic*

[2]*Shanghai Institute of Ceramics, Chinese Academy of sciences,
585 He-shuo Road, Shanghai 201899, P.R. China*

**Abstract**

Electron and hole trapping was studied in the $Ce^{3+}$ and $Pr^{3+}$ doped $Lu_2Si_2O_7$ scintillation single crystals (LPS:Ce and LPS:Pr) by Electron Paramagnetic Resonance (EPR). Detailed EPR measurements of the X-ray irradiated LPS crystals revealed that holes generated by irradiation are predominantly trapped at oxygen lattice ions creating $O^-$ centers. The same X-ray irradiation creates also electron type centers which were attributed to $Lu^{2+}$ ions, where the trapped electron at Lu lattice ion is stabilized by a defect nearby, such as oxygen vacancy and $Ir^{3+}$ impurity ion. Both the hole and electron centers can be thus considered as a bound small polarons, which makes the charge trapping in scintillation mechanism quite competitive. The hole $O^-$ and electron $Lu^{2+}$ centers show thermal stability well above room temperature. The thermal decays of their concentrations correlate well with the appearance of the thermally stimulated luminescence glow peaks at 470-550 K. The presence of the same intrinsic traps in the Ce and Pr doped LPS crystals suggests that difference in the light yield of these crystals is an intrinsic property of the $Ce^{3+}$ and $Pr^{3+}$ activator centers in LPS lattice. Charge traps origin in this pyrosilicate structure and their role in scintillation mechanism is compared with the results previously described in literature in orthosilicates.

**1. Introduction**

Along with the Ce doped lutetium oxyorthosilicate $Lu_2SiO_5$ (LSO:Ce), the Ce and Pr doped lutetium pyrosilicate $Lu_2Si_2O_7$ (LPS) crystals are very promising candidates for detection of γ rays in positron emission tomography (PFT) and high energy calorimetry [1]. LSO:Ce and LYSO:Ce crystals are currently used in scintillation detectors in PET scanners and various codoping schemes have been reported in last decade to improve their performance further [2,3,4,5]. In this respect, LPS:Ce shows even better scintillation properties. In particular, having approximately the same light yield (LY), energy resolution, and decay time as reported in LSO:Ce, it is free from an intense afterglow [6,7] usually present in LSO:Ce. However, the experimentally measured LY of 26000-30000 ph/MeV is still much smaller than the theoretical maximum LY of about 55 000-60 000 ph/ MeV [2,8]. The difference between theoretical and experimental light outputs could be linked either to defects present in the material that could trap electrons or holes created during irradiation, or to efficiency of the electron–hole recombination at $Ce^{3+}$ or $Pr^{3+}$ centers. Several peaks observed around 460 and 550 K in the thermally luminescence (TL) curve [6,7,9] confirm the existence of traps in the material.

As emphasized in [10], serious disadvantage of this crystal is a lack of light yield reproducibility for crystals grown by the Czochralski method. It was suggested [10] that some impurities or defect states (traps) are responsible for the light yield quenching as the $Ce^{3+}$ in this lattice does not show any significant intrinsic luminescence quenching around room temperature. Electron paramagnetic resonance (EPR) study has shown that $Ir^{3+}$ ions penetrated in the crystal from Ir crucible could be responsible for the LY quenching [10]. Annealing treatments under different temperatures and different atmospheres suggest a significant influence of oxygen vacancies on the light output value as well [7].

More complicated situation takes place with $Pr^{3+}$ doped LPS. As a rule, the high energy shifted 5d-4f transition of this ion provides faster response compared to the $Ce^{3+}$. Therefore, LPS:Pr also attracted attention as a perspective fast scintillation material. But as compared to LPS:Ce, it shows much lower LY [11]. It was explained as due to the intense slow 4f-4f emission resulting from efficient energy transfer from self-trapped excitons to 4f states of the $Pr^{3+}$ [11,12]. Later, the radioluminescence (RL) measurement of LPS:Pr powder prepared by sol-gel method did not show any significant contribution of the 4f-4f emission [13]. Furthermore, some of Czochralski grown LPS:Pr crystals demonstrated the radioluminescence 5d-4f emission of $Pr^{3+}$ comparable in amplitude with $Ce^{3+}$ doped crystals [14]. More detailed study of this phenomenon [15] shown that the RL efficiency increase in LPS:Pr crystals is due to the progressive filling of deep traps responsible for the TL peaks at 460 and 515 K, which compete with $Pr^{3+}$ ions in free carriers capturing during irradiation. After filling of the traps the RL increases up to a factor of 14. Because the actual origin of a trap cannot be determine by TL itself, it was speculated that the revealed traps in LPS are of intrinsic type, most probably of a oxygen-vacancy type.

All the mentioned facts demonstrate that the scintillation properties of LPS are quite sensitive to defects created in crystal during growth process which origin must be determined in order to further improve scintillation efficiency of LPS crystals. As the expected defects are intrinsic ones, their local structure and mechanism of creation are of interest for chemistry of materials as well.

Charge traps in a scintillating material can be usually monitored by measurements of TL glow peaks after the material was exposed to ionizing irradiation [16]. However, this method does not reveal determination of actual nature of a trapping center. In this respect, in addition to optical characterization the advanced EPR methods can provide a microscopic insight into the defects creation and their structure. In particular, EPR allows to distinguish accompanying electron and hole localized states which can affect the efficiency and time characteristics of scintillation mechanism in the Ce or Pr-doped scintillation crystals.

In this paper we present the results of detailed EPR study of the Ce and Pr doped LPS crystals grown by Czochralski method. It was found that, independently on activator ion type, the same trapping centers are presented in the both crystals: holes generated by irradiation are trapped at oxygen lattice ions and electrons are trapped at lutetium ions. The trapped holes and electrons are stabilized at lattice ions by a defect nearby. The obtained data suggest that these intrinsic traps in LPS are of polaronic type. This explains their highly competing role in the scintillation process.

## 2. Experimental methods

The crystals were grown from melt by Czochralski method in atmosphere of high pure nitrogen by using iridium crucibles [7]. The initial praseodymium and cerium concentrations in the melt were 0.5 and 0.3 at. %, respectively, with respect to total rare earth sites. It is worth mentioning that the same crystals have been already characterized by TL and other optical measurements [7,9].

LPS crystallizes in the monoclinic structure in the space group C2/m [17]. This lattice exhibits a single crystallographic site for $Lu^{3+}$ ions with six oxygen neighbors forming a distorted octahedron with the $C_2$ point symmetry. For EPR measurements, crystals were cut in three orthogonal planes ($a*b$), ($bc$), and ($a*c$). The axis $a*$ was deflected from the crystallographic axis $a$ by an angle of 12° in order to satisfy the orthogonality between crystal planes that was necessary for determination of g factors from EPR spectra.

EPR measurements were performed in the X-band (9.4 GHz) with a commercial Bruker EMX spectrometer at the temperatures 10-100 K. An X-ray tube operating at a voltage and current of 55 kV and 30 mA, respectively, with Co anode (ISO-DEBYEFLEX 3003 Seifert Gmbh.) was used as a source of X-ray irradiation of crystals. The irradiation dose was about 1 kGray.

## 3. Results and discussion
### 3.1. EPR spectra in x-ray irradiated crystals

As grown LPS crystals, besides of the $Ce^{3+}$ EPR spectra ($Pr^{3+}$ EPR transitions are not visible in the X-band), contained spectra of many other rare earth ions, such as $Yb^{3+}$, $Er^{3+}$, $Dy^{3+}$. Characteristics of these paramagnetic impurity ions will be a subject of separate publication. In the present study we are interested in the intrinsic centers related to trapped hole and electron localized states. As an example, Fig. 1 shows EPR spectrum measured in LPS:Pr before (a) and after (b) X-ray irradiation at room temperature. Before irradiation, there is only one spectrum (designated as Lu1) related to electron traps with well resolved hyperfine (HF) structure. As it will be shown below this HF structure originates from the interaction of an electron spin with the nuclear spin of $^{175}Lu$ isotope (nuclear spin I = 7/2, 100% natural abundance). After X-ray irradiation at room temperature, two new spectra appear: one spectrum at magnetic fields 390-400 mT (designated as trapped electron Lu2 center), and the second one at 332-335 mT. Both spectra show complex HF structure. The spectrum in the lower magnetic fields is assigned to a hole trapped at oxygen ion, i.e. to $O^-$ center (see, Section 3.1.2).

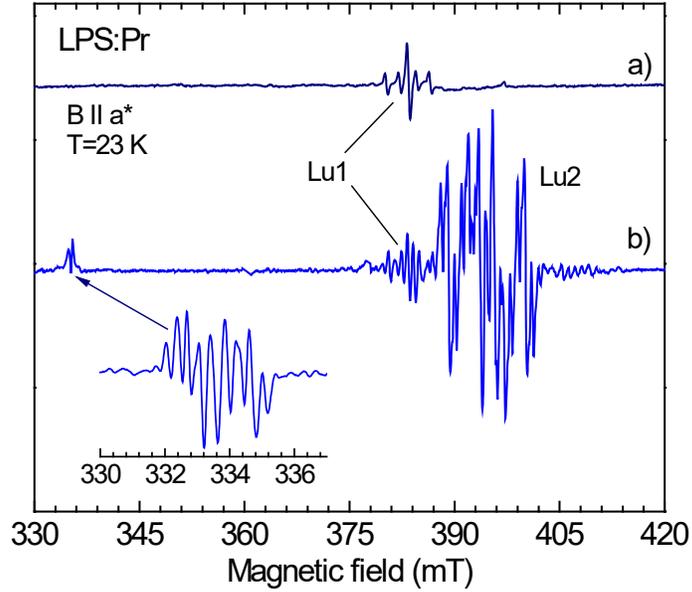

Fig. 1. EPR spectrum before (a) and after (b) X-ray irradiation of LPS:Pr crystal at room temperature. Inset shows spectrum of hole-trapped center recorded at 100 K, when the spectrum is not distorted by spin-lattice relaxation effects as at 23 K.

*3.1.1. Trapped electron centers*

Angular dependencies of resonance fields (center of gravity of the HF structure) of the Lu1 and Lu2 spectra demonstrate very similar behavior (Fig. 2). For the Lu1 center they were fitted by the following $g$ factors: $g_1 = 1.746(1)$, $g_2 = 1.813(1)$, $g_3 = 1.957(1)$, where the three principal axes (1, 2 and 3) coincide with the crystallographic a*, b, and c axes. The obtained $g$ factors are typical for $d^1$ ions [18].

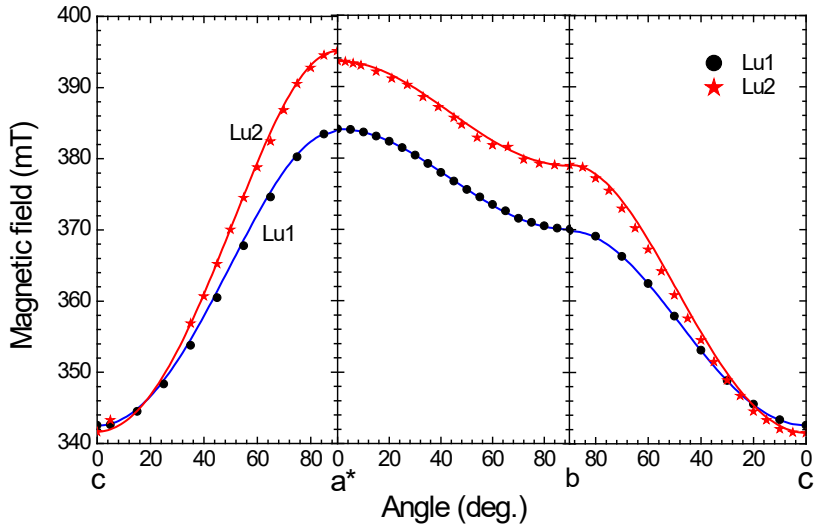

Fig. 2. Angular dependence of the Lu1 and Lu2 resonance fields (center of gravity of their HF structure) measured in three orthogonal planes.

In order to clarify origin of both these centers, we firstly analyzed HF structure of the Lu1 center. Its HF structure consist of eight equidistant components at magnetic field direction along the *c* crystal axis (Fig. 3), suggesting that it originates from an isotope with nuclear spin 7/2 and abundance around 100%. However, as the magnetic field deviates from the *c* axis, the HF structure becomes unusually complex suggesting that there is a strong contribution of the nuclear quadrupole interaction. It was confirmed by simulation of the HF structure with using the following spin Hamiltonian:

$$H = \mu_B \mathbf{B}\mathbf{g}\mathbf{S} + \mu_n \mathbf{B} g_n \mathbf{I} + \mathbf{S}\mathbf{A}\mathbf{I} + \frac{\nu_Q}{2}\left[I_z^2 - \frac{1}{3}I(I+1)\right], \qquad (1)$$

where $\nu_Q = \dfrac{3e^2 qQ}{h 2I(2I-1)}$ is the quadrupole frequency. Here $\mu_B$ and $\mu_n$ are Bohr and nuclear magnetons, respectively; **B** is magnetic field vector, $g_n$ is nuclear g factor, **A** is HF tensor, and $I_z$ is component of nuclear spin along the main axis of electric field gradient (EFG), which is assumed to be aligned along the *c* crystal axis. This assumption is supported by the fact that at the *c* crystal direction the HF structure is not influenced by quadrupole effects. Because forbidden transitions are dominating in HF structure at most magnetic field directions, the spin Hamiltonian (1) was solved numerically to provide correct calculation of resonance fields and probabilities of transitions in electron and nuclear spin systems [19]. The following HF parameters were obtained from the simulation: A = $7.0 \times 10^{-4}$ cm$^{-1}$ (or 21 MHz), $\nu_Q \approx 100$ MHz, i.e. $\nu_Q \gg$ A. The HF constant is nearly isotropic; quadrupole transitions are seen in the HF structure at angles close to the a* axis direction as a small intensity satellites (fig. 3b). The calculated angular dependence of the HF components including their intensities adequately describe all characteristic features of the measured ones (Fig. 3b). In particular, eight HF lines group into two structures divided by enough large distance about 3.5 mT at angles between 20–70$^0$.

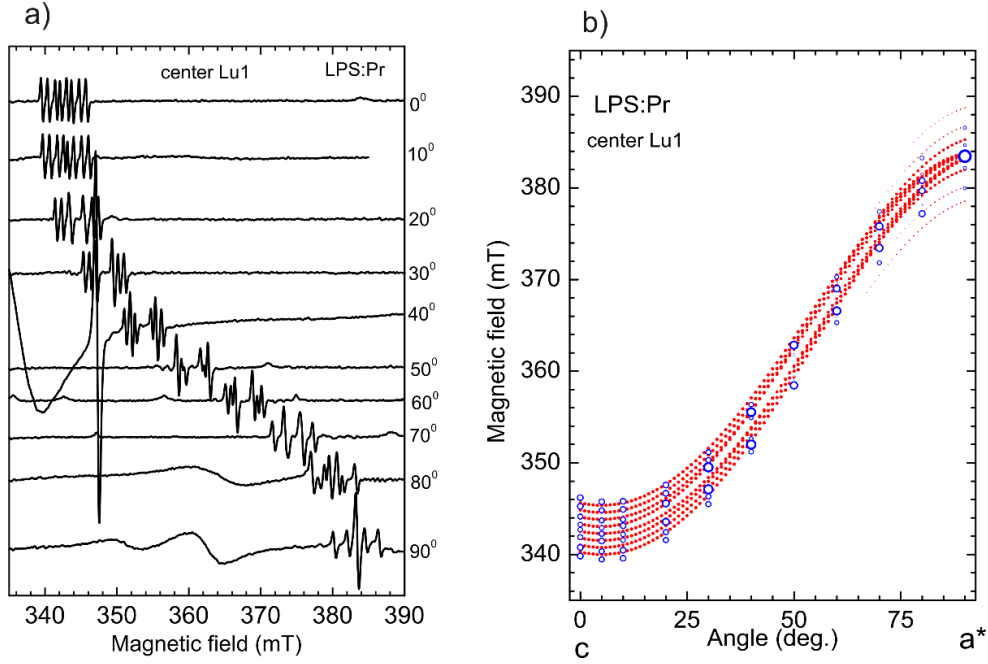

Fig. 3. (a) Angle variation of the Lu1 spectrum in ca* plane and (b) corresponding numerical fit of the HF lines position. In graph (b) red dots are calculated resonance fields with the step of 1 deg., blue cycles are experimental fields destructed from the spectra in graph (a). Size of the cycles and dots are proportional to intensity of spectral component.

Taking into account the *g* factor values, nuclear spin, and huge quadrupole frequency for the center Lu1, the most probable candidate for paramagnetic ion can be $Lu^{2+}$, which has the $5d^1$ outer electronic shell, $^{175}Lu$ isotope with nuclear spin 7/2, natural abundance 97.4%, and huge quadrupole moment eQ = 349 fm$^2$. Among other $nd^1$ ions, only $Ta^{4+}$ has suitable isotope with comparable quadrupole moment. Because such specific HF structure is observed also for other X-ray created paramagnetic defects in our crystals (Lu2, O⁻), we prefer a model with Lu ion, which is the lattice cation. More arguments in favor of the $Lu^{2+}$ - related center will be given below.

*g* factors of the Lu2 center are close to those of the Lu1 center: $g_1$ = 1.700(1), $g_2$ = 1.770(1), $g_3$ = 1.962(1). Within the *g* factors determination error they coincide with the *g* factors of the Ir-related center described in [10], where it appeared also after X-ray irradiation. It was assigned to $Ir^{4+}$ ion in a low-spin state of this ion. In our opinion, the similarities in the symmetry and *g* factors values of the both Lu1 and Lu2 centers suggest the same origin of these centers, namely that they belong to $Lu^{2+}$ ions, which are perturbed by a defect in a different way.

In spite that the Lu2 center has much more complex HF structure, both Lu centers demonstrate similarity in angle dependence of their HF structures as it can be seen from spectra presented in Fig. 4, where the spectral lines of both centers are visible. Namely, at most directions of magnetic field, the HF lines are grouped into two structures related to dominating quadrupole interactions. The HF structure for Lu2 center is very rich at B || b (bottom spectrum in Fig. 4). Moreover, it contains forbidden quadrupole transitions visible well as satellites at 371 and 387 mT.

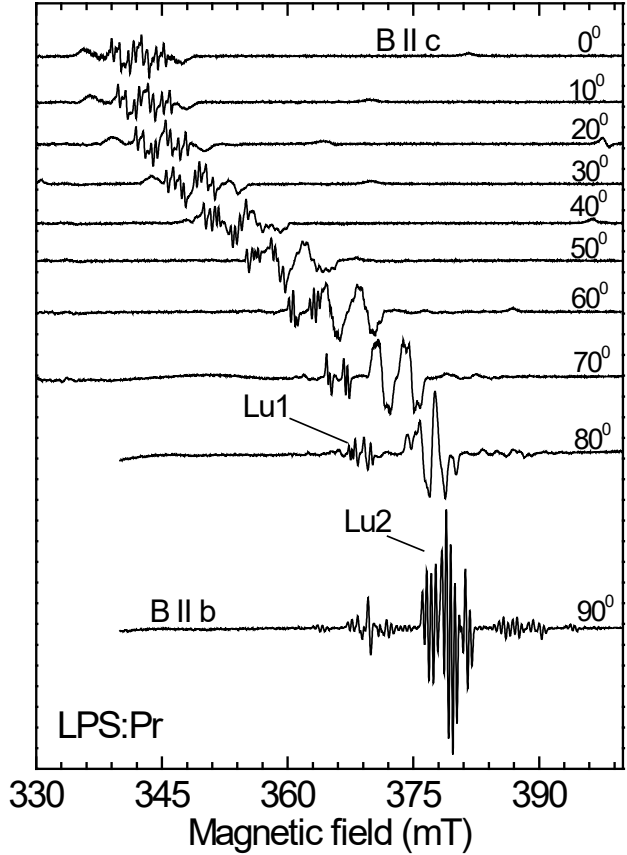

Fig. 4. Angle variation of the Lu2 spectrum in the *bc* plane. These spectra contain also small intensity spectral lines from the Lu1 center.

The quadrupole effects for Lu2 center must diminish at B || c similar as for Lu1 center. We performed simulation of the Lu2 HF structure at B || c for LPS crystal doped with Ce and annealed in air. This annealing decreases concentration of the Lu1 centers and their contribution to the total spectrum at B || c is negligibly small. In this way, we wanted to check if the largest HF splitting is produced by two Ir isotopes as it was concluded in Ref. 10. The result of the simulation in shown in Fig. 5 where the absorption spectrum was used instead of its derivative to distinctly show big disagreement between the measured and calculated HF structures. In particular, the calculated HF pattern consists of four components of equal intensity as contributions from two Ir isotopes are not distinguished separately. The measured HF structure can be fitted by, at least, six non-equidistant HF components with separation between components much larger than that predicted by difference in the nuclear *g* factors of two Ir isotopes. Consequently, this HF structure can

hardly belong to Ir isotopes. It is created, as in case of the Lu1 center, by $^{175}$Lu isotope with strong contribution of quadrupole interaction.

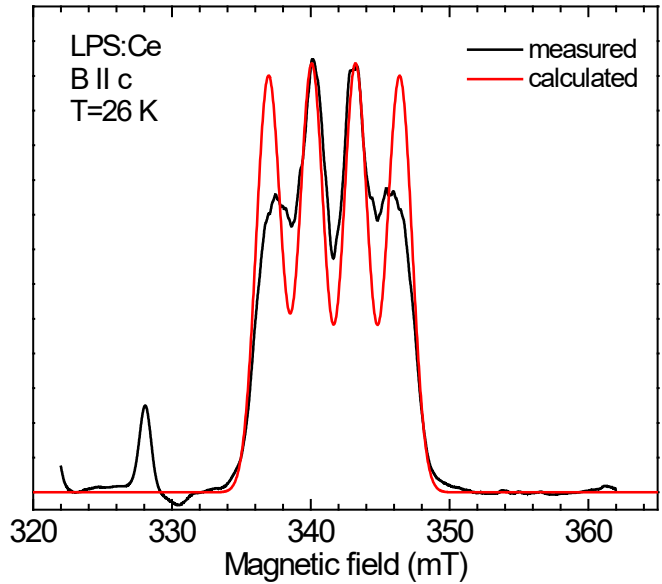

Fig. 5. Simulation of the Lu2 HF structure assuming that only $^{191}$Ir and $^{193}$Ir isotopes contribute to the HF structure. Dark line presents experimental absorption spectrum; red line is simulated spectrum.

### 3.1.2. Trapped hole O⁻ center

As it was mentioned above, besides of electron type centers, X-ray irradiation creates also a hole type center. Angular dependencies of resonance fields (center of gravity of HF structure) of this hole center in LPS:Pr (Fig. 6) were fitted by the following *g* factors: $g_1 = 2.007(1)$, $g_2 = 2.011(1)$, $g_3 = 2.024(1)$, where three principal axes (1, 2 and 3) coincide with the crystallographic axes a*, b, and c. These *g* factor values are typical for O⁻ hole center usually created by ionizing radiation in oxide materials [16,20,21,22]. Therefore, it was reasonable to assign this spectrum to a hole trapped at oxygen ion (O⁻, $2p^5$, S=1/2). Similar O⁻ center is created by X-ray irradiation in LPS:Ce crystal as well. Its spectrum shows approximately the same *g* factors. But, due to much broader HS structure (Fig. 7, bottom spectrum) *g* factor actual values were not determined.

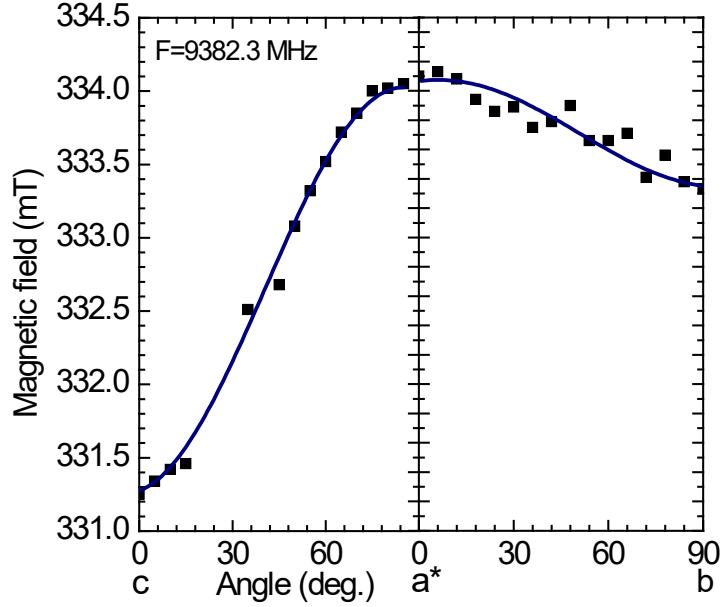

Fig. 6. Angular dependencies of O⁻ resonance fields (center of gravity of the HF structure) measured for two rotation planes (ca* and a*b) in LPS:Pr crystal.

O⁻ spectrum has a complex HF structure which can hardly be analyzed. As an example, behavior of the HF structure with the angle under rotation of the LPS:Pr crystal in the (ca*) plane is shown in Fig. 7. One can see, however, that there is a similarity in the HF pattern of this center and the Lu ones (namely, the HF lines are again grouped into two structures) suggesting that here the unusual HF structure originates also from $^{175}$Lu isotopes with strong manifestation of quadrupole effects on HF splitting and appearance of strong forbidden transitions. The lowermost spectrum in Fig. 7 demonstrates HF structure of the O⁻ center in LPS:Ce. At this given crystal orientation B ∥ [110], the HF pattern contains 15 equidistant lines, which can be attributed to interaction of "hole" spin with nuclear spins of two $^{175}$Lu isotopes (the number of HF lines is calculated as $n = 2I_{Lu} + 1 = 15$). In LPS lattice, there are two oxygen ions sites, O2 and O3, which are directly linked to Lu ions in the first neighbor shell. Namely, the O2 ion has two Lu ions at the equal distance of 2.209 Å [17]. Therefore, one can suppose that hole is located at the O2 ion. Because the HF pattern in LPS:Pr is almost two times narrower (it contains less spectral lines than that in LPS:Ce) one of the two Lu ions in the O⁻ center in LPS:Pr is substituted by an impurity ion (or this center includes Lu vacancy) which, as it will be shown below, increases also thermal stability of this center.

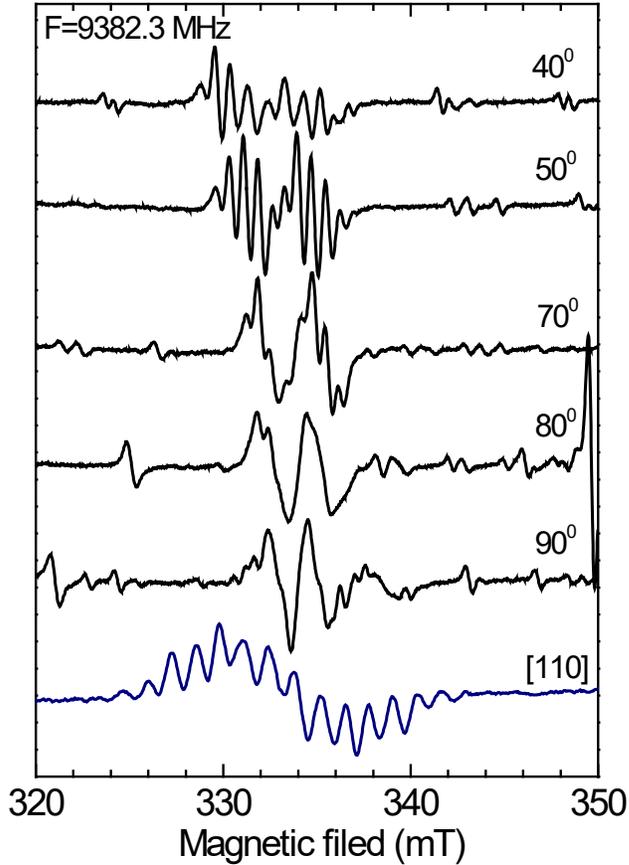

Fig. 7. Angle variation of the O⁻ center HF structure measured in the ca* plane of LPS:Pr. The HF structure is badly resolved at angles < 40⁰ as it is overlapped with stronger lines from other centers. The bottom spectrum presents HF structure of the O⁻ center in LPS:Ce.

### 3.2. Temperature stability of the electron and hole centers and comparison with TSL

Firstly we checked how the high-temperature annealing in air atmosphere influences the centers visible in EPR. Fig. 8 compares spectra in LPS:Ce crystal before and after annealing in air at 1400 ⁰C for 6 hours. One can see that the high-temperature annealing completely "erases" all the X-ray created spectra. X-ray irradiation of the annealed crystal restores the O⁻ and Lu2 centers almost with the same EPR intensity, while the Lu1 center restores with much lower EPR intensity indicating that concentration of these centers substantially decreased after annealing. Such behavior of EPR intensity usually suggests that center contains oxygen vacancy ($V_O$). Trapped electron is thus located at Lu ion in the vicinity of an oxygen vacancy. EPR intensity of the Lu2 center only slightly decreases in annealed crystal suggesting that this center is not related to an oxygen vacancy. Trapped electron at the Lu ion in this center is thus stabilized by another defect. Because the Lu2 center is present only in crystals grown in Ir crucible [10] one can conclude that this perturbing defect is an $Ir^{3+}$ impurity ion.

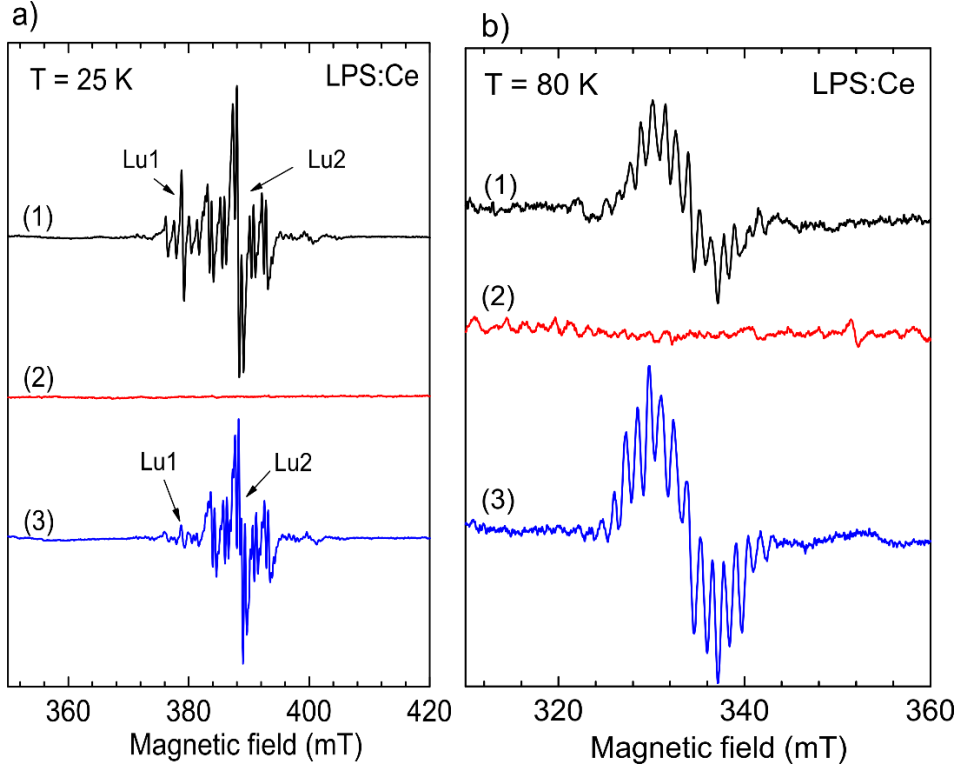

Fig. 8. EPR spectra of the Lu1 and Lu2 centers (a) and O$^-$ center (b) in LPS:Ce crystal measured (1) in X-ray irradiated "as grown" crystal, (2) in crystal annealed in air at 1400 $^0$C, (3) in crystal annealed in air at 1400 $^0$C and then X-ray irradiated.

Further experiments were devoted to determination of thermal stability of the X-ray induced centers in both LPS:Pr and LPS:Ce crystals. In this "pulsed annealing" experiments, after X-ray irradiation at room temperature, the sample was heated to a certain temperature $T_{ann}$, held at that temperature for 5 - 6 minutes, and then quickly cooled down to room temperature. After that EPR spectrum was measured either at 80-110 K for O$^-$ center or at 25 K for Lu centers.

The thermal stability data for both crystals are presented in Fig. 9. One can see that the Lu1 center in both crystals is stable at least up to 550 K. On the contrary, the O$^-$ center in LPS:Ce shows the smallest stability, only to about 400 K. In LPS:Pr, the O$^-$ center is stable to about 420 K. In this crystal, thermally liberated holes from oxygen ions recombine at Lu2 center markedly decreasing concentration of the Lu2 centers. Lu2 concentration further decreases with temperature increase. If we compare data in Fig. 9a with TSL glow peaks measured in the similar LPS:Pr crystals (Fig.7 of Ref. 9) one can notice that position of the main TSL peak at 470 K correlates well with the temperature where O$^-$ centers disappear ( the TSL peak is slightly shifted to higher temperature due to faster heating in TSL measurements as compared to our pulse annealing condition). The high-temperature shoulder in the thermal decay of the Lu2 concentration suggests that electrons are also released at temperatures 450-550 K providing their preferable recombination with holes at $Pr^{3+}$ activator ions. Corresponding high-temperature shoulder is seen also in TSL [9].

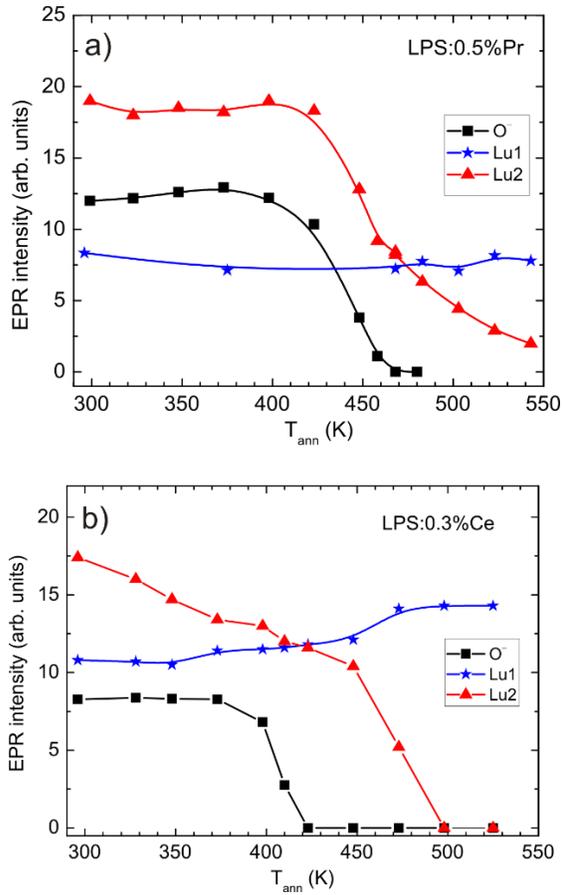

Fig. 9. Dependence of the $O^-$, Lu1 and Lu2 EPR intensity (concentration of the ions) on pulsed annealing temperature in the X-ray irradiated LPS:Pr (a) and LPS:Ce (b) crystal. The presented EPR intensities do not reflect the actual ratio of the ion concentrations.

X-ray created electron and hole centers in LPS:Ce crystal show similar temperature intervals of stability. However, their actual behavior with annealing temperature differs from that in LPS:Pr. One can notice that thermal destruction of $O^-$ centers at ≈ 400 K does not lead to visible decrease in Lu centers concentration (Fig. 9b). Any TSL peak is not seen at these temperatures as well [23]. The main TSL peak is located at 475 K, where Lu2 centers decays in concentration. This fact suggests that liberated holes do not recombine at the Lu centers but are rather retrapped by $Ce^{3+}$ ions forming $Ce^{4+}$ states. At further stage, electron thermally excited from a Lu2 center recombines with hole at $Ce^{4+}$ ion providing 5d – 4f emission visible in TSL. The Lu2 centers also show gradual decrease of their concentration with temperature increase. It indicates an electron transfer from Lu2 centers to other defects or impurity ions via tunneling mechanism. This process is so strong that the Lu2 spectrum becomes almost invisible after two weeks storage of the crystal at room temperature. Firstly, there is charge transfer from the Lu2 centers to Lu1 ones as the Lu1 concentration gradually increases with temperature increase. We cannot also exclude the charge transfer to $Ce^{3+}$ ions. For instance, the tunneling-governed emission was reported for LPS:Pr crystals [15].

## 4. Discussion and concluding remarks

It should be noted that in previous publications, the $Ir^{3+}$ impurity ions were considered as traps for holes in LPS:Ce crystals [6,23]. Our study shows that as in many other oxide materials, namely oxygen lattice ions serve as effective trap sites for holes [20,21,22]. On the contrary, we found that the $Ir^{3+}$ ions serve as a stabilizing defect for electron trapping at Lu lattice ions. Both the electron and hole trapping phenomena point to polaronic trapping mechanism. The polaronic effect in LPS lattice is much stronger than that in LSO (YSO) lattice where both the electrons and holes created by irradiation are coupled to lattice ions with much smaller energy. Corresponding $O^-$ and $Y^{2+}$ - $V_O$ (or $Lu^{2+}$ -$V_O$) centers are thermally stable only below room temperature [24]. Similar centers in LPS are stable up to 400-500 K. Such large difference in the thermal stability of hole and electron centers in LPS and LSO lattices is related to the fact that in LSO electron or hole is trapped at Si-unbound oxygen ion or vacancy created at this oxygen site. This oxygen ion in LSO is only weakly coupled to lattice providing weak bounding energy for trapped hole or electron [25].

Low-temperature X-ray irradiation of LPS:Pr and LPS:Ce crystals at 77 K creates the same $O^-$ and $Lu^{2+}$ centers similarly as the irradiation at RT. In spite that both these crystals contain TL glow peaks at 102 K, 211 K, and 265 K [23], any additional EPR spectra do not appear at X-ray irradiation at 77 K indicating that the mentioned above TL peaks visible below room temperature are created by paramagnetic silent defects.

In many $Ce^{3+}$ and $Pr^{3+}$ doped scintillation crystals, the 4+ valence state of these activator ions plays an important role in scintillation mechanism [26,27,28,29]. Both these ions can also act as effective traps for holes. They store holes before their radiative recombination with electrons. For instance, no $O^-$ hole centers were detected in LSO:Ce crystals [24]. On the contrary, in the Ce and Pr doped LPS crystals namely the $O^-$ centers appear after irradiation indicating that primarily holes are trapped at oxygen ions. Only after thermal destruction of the $O^-$ centers the holes can be retrapped at the $Ce^{3+}$ or $Pr^{3+}$ ions or, alternatively, the effective trapping of holes by activator ions would be possible only after complete filling of the oxygen ions trapping sites, so that these oxygen centers would not more compete with the Ce trapping. This explains the marked increase of the radioluminescence after prolonged irradiation of LPS crystals [15].

It is difficult to predict if the $Ca^{2+}$ or $Mg^{2+}$ cooping of LPS will have positive effect on scintillation light yield by stimulating the $Ce^{3+}$ to $Ce^{4+}$ recharge like in LSO [2] as these divalent ions can stabilize further $O^-$ centers. However, a positive effect from the divalent ions codoping could be expected after filling of oxygen trapping sites. Actual concentration of the Ce ions is also a critical aspect in the improving the LPS scintillation efficiency. This concentration has to be small enough to suppress charge transfer from defect states via tunneling mechanism. At the same time, it has to be large enough to provide effective energy transfer from the host and electron-hole radiative recombination at Ce ions.

Finally, note since with both the Ce and Pr doping in LPS host there are the same traps and impurities (at least those detectable by EPR) the difference in the light yield in these crystals is hardly related to traps or impurities in the crystals. It is rather an intrinsic property of the $Ce^{3+}$ and $Pr^{3+}$ activator centers and their interplay with the LPS lattice.

**Acknowledgements:** The financial supports of the Czech Science Foundation grant 17-09933S and the Ministry of Education, Youth and Sports of Czech Republic, projects SAFMAT LO1409 and CZ.02.1.01/0.0/16_013/0001406 are gratefully acknowledged.

**References**

[1] C. W. E. van Eijk, Inorganic scintillators in medical imaging. Phys. Med. Biol. **47,** R85-R106 (2002).

[2] K. Yang, Ch. L. Melcher, P. D. Rack, L. A. Eriksson, Effects of Calcium Codoping on Charge Traps in LSO:Ce Crystals. IEEE Trans.Nucl.Science **56**, 2960-2965 (2009).

[3] S. Blahuta, A. Bessière, B. Viana, P. Dorenbos, and V. Ouspenski, Evidence and Consequences of Ce in LYSO:Ce,Ca and LYSO:Ce,Mg Single Crystals for Medical Imaging Applications. IEEE Trans. Nucl.Science **60**, 3134-3141 (2013).

[4] Y. Wu, M. Koschan, C. Foster, Charles L. Melcher, Czochralski Growth, Optical, Scintillation, and Defect Properties of $Cu^{2+}$ Codoped $Lu_2SiO_5$:$Ce^{3+}$ Single Crystals. Cryst. Growth Des. **19**, 4081−4089 (2019).

[5] Y. Wu, J. Peng, D. Rutstrom, M. Koschan, C. Foster, Ch. L. Melcher, Unraveling the Critical Role of Site Occupancy of Lithium Codopants in $Lu_2SiO_5$:$Ce^{3+}$ Single-Crystalline Scintillators. ACS Appl. Mater. Interfaces **11**, 8194−8201 (2019).

[6] L. Pidol, A. Kahn-Harari, B. Vianan, B. Ferrand, P. Dorenbos, J. T. M. de Hass, C. W. E. van Eijk, and E. Virey, Properties of $Lu_2Si_2O_7$:$Ce^{3+}$, a fast and efficient scintillator crystal. J. Phys.: Condens. Matter **15**, 2091-2102 (2003).

[7] F. He, R. Guohao, W. Yuntao, X. Jun, Y. Qiuhong, X Jianjun, C. Mitch, C. Chenlong, Optical and thermoluminescence properties of $Lu_2Si_2O_7$:Pr single crystal. Journal of rare earths **30**, 775-779 (2012).

[8] P. Dorenbos, Fundamental Limitations in the Performance of $Ce^{3+}$, $Pr^{3+}$, and $Eu^{2+}$ -Activated Scintillators. IEEE Trans.Nucl.Science **57**, 1162-1167 (2010).

[9] E. Mihóková, M. Fasoli, F. Moretti, M. Nikl, V. Jary, G. Ren, A. Vedda, Defect states in $Pr^{3+}$ doped lutetium pyrosilicate. Optical Materials **34**, 872–877 (2012).

[10] L. Pidol, O. Guillot-Noel, M. Jourdier, A. Kahn-Harari, B. Ferrand, P. Dorenbos, D. Gourier, Scintillation quenching by $Ir^{3+}$ impurity in cerium doped lutetium pyrosilicate crystals. J. Phys.: Condens. Matter. **15**, 7815-7821 (2003).

[11] L. Pidol, B. Vianan, A. Kahn-Harari, A. Bessiere, P. Dorenbos, Luminescence properties and scintillation mechanisms of $Ce^{3+}$-, $Pr^{3+}$- and $Nd^{3+}$-doped lutetium pyrosilicate. Nucl. Instrum. Methods Phys. Res. A **537**, 125-129 (2005).

[12] L. Pidol, B. Vianan, A. Caltayries, P. Dorenbos, Energy levels of lanthanide ions in a $Lu_2Si_2O_7$ host. Phys. Rev. B **72**, 125110 (2005).

[13] M. Nikl, A. M. Begnamini, V. Jary, D. Niznansky, E. Mihokova, $Pr^{3+}$ luminescence center in $Lu_2Si_2O_7$ host. Phys. Stat. Sol. RRL **3**, 293-295 (2009).

[14] M. Nikl, G. Ren, D. Ding, E. Mihokova, V. Jary, H. Feng, Luminescence and scintillation kinetics of the $Pr^{3+}$ doped $Lu_2Si_2O_7$ single crystal. Chemical Physics Letters **493**, 72–75 (2010).

[15] E. Dell'Orto, M. Fasoli, G. Ren, A Vedda, Defect-Driven Radioluminescence Sensitization in Scintillators: The Case of $Lu_2Si_2O_7$:Pr. J. Phys. Chem. C **117**, 20201−20208 (2013).

[16] M. Nikl, V. V. Laguta and A. Vedda, Complex oxide scintillators: Material defects and scintillation performance. Phys. Status Solidi B **245**, 1701–1722 (2008).

[17] F. Bretheau-Raynal, M. Lance, P. Charpin, Crystal data for $Lu_2Si_2O_7$. J. Appl. Cryst. **14**, 349-350 (1981).

[18] A. Abragam, B. Bleaney, Electron Paramagnetic Resonance of Transition Ions (Clarendon Press, Oxford (1970)).

[19] EPR spectra and angular dependencies were simulated using the "Visual EPR" programs by V. Grachev (www.visual-epr.com).

[20] V. V. Laguta and M. Nikl, Electron spin resonance of paramagnetic defects and related charge carrier traps in complex oxide scintillators. Phys. Stat. Sol. B **250**, 254-260 (2013).

[21] O. F. Schirmer, $O^-$ bound small polarons in oxide materials J. Phys.: Condens. Matter **18**, R667 (2006).

[22] V. Laguta, M. Buryi, J. Pejchal, V. Babin and M. Nikl, Hole Self-Trapping in $Y_3Al_5O_{12}$ and $Lu_3Al_5O_{12}$ Garnet Crystals. Phys. Rev. Appl. **10**, 034058 (2018).

[23] H. Feng, D. Ding, H. Li, Sh. Pan, X. Chen, G. Ren, Annealing effects on Czochralski grown $Lu_2Si_2O_7$:$Ce^{3+}$ crystals under different atmospheres. J. Appl. Phys. **103**, 083109 (2008).

[24] V.V. Laguta, M. Buryi, J. Rosa et al., Electron and hole traps in yttrium orthosilicate single crystals: The critical role of Si-unbound oxygen. Phys. Rev. B **90**, 064104 (2014).

[25] B. Liu, Z. Qi, M. Gu, X. Liu, S. Huang, C. Ni, First-principles study of oxygen vacancies in $Lu_2SiO_5$. J. Phys.: Condens. Matter **19**, 436215 (2007).

[26] M. Nikl, V. Babin, J. Pejchal, V.V. Laguta et al., The stable $Ce^{4+}$ center: a new tool to optimize Ce-doped oxide scintillators, IEEE Trans. Nucl. Sci. **63**, 433 (2016).

[27] Y. Wu, F. Meng, Q. Li, M. Koschan, and Ch. L. Melcher, Role of $Ce^{4+}$ in the Scintillation Mechanism of Codoped $Gd_3Ga_3Al_2O_{12}$:Ce, Phys. Rev. Appl. **2**, 044009 (2014).

[28] F. Moretti, K. Hovhannesyan, M. Derdzyan, G. A. Bizarri, E. D. Bourret, A. G. Petrosyan, C. Dujardin, Consequences of Ca Codoping in $YAlO_3$:Ce Single Crystals. Chem. Phys. Chem. **18**, 493 – 499 (2017).

[29] J. Pejchal, M. Buryi, V. Babin, P. Prusa, A. Beitlerova, J. Barta, L. Havlak, K. Kamada, A. Yoshikawa, V. Laguta, M. Nikl, Luminescence and scintillation properties of Mg-codoped LuAG:Pr single crystals. J. Lumin. **181**, 277–285 (2017).